\documentclass[aps,prl,twocolumn,preprintnumbers,showpacs,superscriptaddress,groupedaddress]{revtex4}  
\usepackage{graphicx}  
\usepackage{dcolumn}   
\usepackage{bm}        
\usepackage{amssymb}   

\usepackage{hyperref}

\usepackage[usenames,dvipsnames]{color}
\usepackage{graphicx}
\usepackage{amsmath}
\usepackage{amsfonts}
\usepackage{dsfont}
\usepackage{dcolumn}
\usepackage{bm}
\usepackage{slashed}
\usepackage{pstricks}
\usepackage{slashed}
\usepackage{multirow}
\usepackage{array}
\usepackage{rotating}
\usepackage{xcolor}
\usepackage{epstopdf}
\usepackage{fancybox}
\usepackage{fancyvrb}
\usepackage[normalem]{ulem}
\usepackage{color}
\usepackage{mathdots}
\usepackage{mathrsfs}
\usepackage{multirow}
\usepackage{esint}
\allowdisplaybreaks

\usepackage{hyperref}
\usepackage{cleveref}
\crefname{equation}{Eq.}{Eqs.}
\crefname{figure}{Fig.}{Figs.}
\crefname{table}{Table}{Tables}
\crefname{section}{Section}{Sections}

\def\MeV{\rm MeV}
\DeclareGraphicsExtensions{.jpg,.pdf,.mps,.png,}

\def \d{{\rm d}}

\newcommand{\beqn}{\begin{eqnarray}}
\newcommand{\eeqn}{\end{eqnarray}}
\newcommand{\be}{\begin{equation}}
\newcommand{\ee}{\end{equation}}

\begin{document}

\preprint{MPP-2013-321}

\title{Cogenesis in a universe with vanishing $B-L$
within a gauged $U(1)_x$ extension}

\author{Wan-Zhe~Feng\footnote{Email: vicf@mpp.mpg.de}}
\affiliation{Max--Planck--Institut f\"ur Physik (Werner--Heisenberg--Institut),
80805 M\"unchen, Germany}
\author{Pran~Nath\footnote{Email: nath@neu.edu}}
\affiliation{Department of Physics, Northeastern University, Boston, MA 02115, USA}


\begin{abstract}
  We consider a gauged $U(1)_x$  extension of the standard model and of the minimal supersymmetric
  standard model
  where the dark matter fields are charged under $U(1)_x$ and carry lepton number
  while the  standard model fields and fields of the minimal supersymmetric standard model
  are neutral under $U(1)_x$.  We consider leptogenesis in this class of  models with all fundamental interactions
  having no violation of lepton number, and the total $B-L$ in the universe vanishes.
  Such leptogenesis leads to equal and opposite lepton numbers in
  the visible sector and in the dark sector, and thus also produces asymmetric dark matter.
  Part of the lepton numbers generated in the leptonic sector
  subsequently transfer to the baryonic sector via sphaleron interactions.
  The stability of the dark particles is protected by the $U(1)_x$ gauge symmetry.
  A kinetic mixing between the $U(1)_x$ and the $U(1)_Y$ gauge bosons
  allows for dissipation of the symmetric component of dark matter.
  The case when $U(1)_x$ is $U(1)_{B-L}$ is also discussed for the supersymmetric case.
  This case is particularly interesting in that
  we have a gauged $U(1)_{B-L}$ which ensures the conservation of $B-L$
  with an initial condition of a vanishing $B-L$ in the universe.
  Phenomenological implications of the proposed extensions are discussed,
  which include implications for electroweak physics,
  neutrino masses and mixings,
  and lepton flavor changing processes such as $\ell_i \to \ell_j \gamma$.
  We also briefly discuss the direct detection of the dark matter in the model.
\end{abstract}


\maketitle

{\it Sec.I. Introduction:\label{sec1}}
Three of the important puzzles in cosmology relate to the origin of baryon asymmetry in the Universe,
the nature of dark matter and the cosmic coincidence. Thus the visible universe exhibits an excess of baryons
over anti-baryons and this excess
is often displayed as the baryon number density to the entropy density ratio~\cite{wmap}
\be
B/s\sim 6\times 10^{-10}\,.
\label{0.1}
\ee
The basic tenets of how to generate baryon (lepton)
excess has been known since the work of Sakharov~\cite{Sakharov:1967dj}, and consist of three conditions, i.e.,
the existence of  baryon (or lepton) number violation, the presence of C and CP violating interactions, and
out of equilibrium processes. In the standard model the ratio $B/{s}$ is computed to be
too small to fit observation pointing to the existence of beyond the standard model physics.
Standard model also does not provide us with a candidate for dark matter and
the astrophysical evidence for its presence again points to the existence of new
physics beyond the standard model. Additionally one has the cosmic coincidence puzzle, i.e., the
fact that the amount of dark matter and the amount of visible matter in the Universe are comparable.
Specifically one has~\cite{Ade:2013ktc}
\be
\frac{\Omega_{\rm DM} h_0^2}{ \Omega_{\rm B} h_0^2} \approx 5.5\,.
\label{0.2}
\ee
The comparable sizes of the amounts  of
dark matter and of visible matter point
to the possibility of a common origin of the two.
This can be explained by the so-called asymmetric dark matter hypothesis
where the dark particles are in thermal equilibrium with
the standard model particles or with the particles of the minimal supersymmetric standard model
in the early universe, and thus their chemical potentials
are of the same order.
The satisfaction of~\cref{0.2} then occurs via a constraint on the dark matter mass
(for a sample of recent works see~\cite{AsyDM,Feng:2012jn}
 and for reviews see \cite{review}).
Alternative schemes where dark matter carrying a lepton number
(or a baryon number) is created first and
a portion of it subsequently transfer to the visible sector have been considered in~\cite{darkogenesis,Feng:2013wn}.
Cogenesis of baryon/lepton asymmetry and the asymmetric dark matter
have also been discussed recently in~\cite{cogenesis}.

An important constraint on model building  is the requirement that dark matter be stable, i.e.,
the dark particles does not decay into lighter standard model particles.
In this work we consider an extension of the standard model and of the minimal supersymmetric standard model
where the dark fields are charged under a $U(1)_x$ gauge symmetry
while the standard model fields are neutral under $U(1)_x$,
which forbids dark particles decay into the standard model particles and thus
guarantees the stability of the dark matter.
Additionally, the asymmetry of the dark particles generated in the early universe
will not be washed out by  Majorana mass terms since they  are forbidden  by the
$U(1)_x$ gauge symmetry.\footnote{
Models which allow a Majorana term for dark matter
can undergo oscillations where the dark particle oscillates to its anti-particle.
Such processes over the lifetime of the universe can produce symmetric dark matter
which can lead to pair annihilation and wipe out the asymmetric dark matter \cite{oscillations}.}
In the supersymmetric case, a gauged $U(1)_{B-L}$ model is also discussed.

Most conventional models of baryogenesis or leptogenesis assume that the fundamental vertices
violate either baryon number or lepton number or both in conformity with the first Sakharov condition~\cite{Fukugita:1986hr,Covi:1996wh}.
However, in this work we consider leptogenesis
where the fundamental interactions conserve lepton number and leptogenesis consists in generating
equal and opposite lepton numbers in the visible and in the dark sectors.
Subsequently the sphaleron processes transmute a part of the leptons into baryons.
The total $B-L$ in the universe is exactly conserved.
This mechanism bypasses the difficulty in the GUT baryogenesis where
a vanishing total $B-L$ implies that
the baryon asymmetry generated would be washed out by the sphaleron interactions.
While this idea has been recently pursued by several authors~\cite{Davoudiasl:2010am,Gu:2010ft},
our analysis differs significantly in structure and in content from previous works~\cite{Davoudiasl:2010am,Gu:2010ft}.\footnote{
Baryogenesis with a gauged $U(1)_B$ symmetry is discussed in~\cite{Perez:2013tea},
where the dark sector and the visible sector carry the opposite baryon number
and the total baryon number in the universe is conserved.
While in this work a pre-existing excess of lepton number has been assumed,
thus the total $B-L$ in the universe is not vanishing.}
A more detailed comparison with these works is given at the end of Sec.4.
Earlier works on Dirac leptogenesis~\cite{DLeptogenesis}
can also generate the asymmetry in the visible sector starting from a $B-L$ vanishing universe.
Due to the tiny Yukawa coupling,
right-handed (Dirac) neutrinos would not be in thermal equilibrium with left-handed neutrinos,
hence the sphaleron interactions which operate only on $SU(2)$ fields,
will not wash out the lepton number stored in the right-handed neutrinos and thus the asymmetry is created.

The outline of the rest of the Letter is as follows: In Sec.2 we discuss
leptogenesis and the generation of asymmetric dark matter
in a non-supersymmetric model where the vertices have no lepton number violation.
The dark matter consists of two fermionic fields which carry the same lepton numbers but opposite $U(1)_x$ charges.
Here we also compute the mass of the dark particles which satisfy the cosmic coincidence of~\cref{0.2}.
In Sec.3 we extend the analysis to the supersymmetric case.
The main difference in the analysis of Sec.3 from the analysis of Sec.2 is that in the supersymmetric case
there are more species of dark matter particles.
Specifically we have four types  of fermionic particles and their bosonic super-partners
which carry different combinations of the lepton numbers and $U(1)_x$ charges.
We also discuss the possibility that $U(1)_x$ is  $U(1)_{B-L}$.
In Sec.4 we discuss the phenomenology related to these models.
Conclusions are given in Sec.5.
\\

{\it Sec.2. Non-supersymmetric Model: }\label{sec2}
 We begin by considering  the  set of fields  $N_i, \psi, \phi, X, X'$
with lepton number assignments $(0, +1, -1, +1/2, +1/2)$.
Here $N_i$ ($i \geq 2$) are Majorana fermions,
$\psi, X,X'$ are Dirac fields and $\phi$ is a complex scalar field.
The fields $N_i, \psi, \phi$ are heavy and will decay into lighter fields
and eventually disappear and there would be no vestige left of them in the current universe.
The dark sector is constituted of two fermionic fields $X,X'$, which
as indicated above each carry a lepton number $+1/2$ and
are oppositely charged under the dark sector gauge group $U(1)_x$ with gauge charges $(+1,-1)$.
All other fields are neutral under $U(1)_x$.  We assume their interactions to have the following form
which conserve both the lepton number and the $U(1)_x$ gauge symmetry:
\begin{equation}
\mathcal{L} =\lambda_i \bar N_i \psi \phi + \beta\, \bar \psi L H
+ \gamma\, \phi \bar X^c X' + h.c.\,,
\label{1.1}
\end{equation}
where the couplings $\lambda_i$ are assumed to be complex and the couplings $\beta,\gamma$ are
assumed to be real. In addition we add mass terms so that
\begin{multline}
- \mathcal{L}_m= M_i \bar N_i N_i +  m_1 \bar \psi \psi + m_2^2 \phi^* \phi \\
  +m_X \bar X X + m_{X'} \bar X' X' \,.
\label{1.2}
\end{multline}
Here $N_i$ have Majorana masses, while $\psi, X, X'$ have Dirac masses.
We assume the mass hierarchy $M_i \gg  m_1 + m_2$, $m_1\sim m_2 \gg m_X+ m_{X'}$.
We will see later that $m_X,m_{X'}$ are around 1~GeV.
Consistent with the above constraint,
$m_1, m_2$ which are the masses of $\psi$ and $\phi$ respectively,
could span a wide range from order of TeV to much higher scales.

In the early universe, the out-of-equilibrium decays of the heavy Majorana fields $N_i$
produce a heavy Dirac field $\psi$ and a heavy complex scalar field $\phi$.
The CP violation due to the complex couplings  $\lambda_i$ generates
an excess of $\psi,\phi$ over their anti-particles $\bar{\psi},\phi^*$
which carry the opposite lepton numbers.
Since the lepton number carried by $\psi$ and $\phi$ always sums up to zero,
the out-of-equilibrium decays of $N_i$ do not generate an excess of lepton number in the universe.
Further, $\psi$ and $\phi$ (as well as their anti-particles) produced in the decay of the Majorana fields $N_i$
will sequentially decay, with $\psi$ (and its anti-particle) decaying into the visible sector fields and $\phi$
(and its anti-particle) decaying into
the dark sector fields.  Their decays thus produce a net lepton asymmetry in the visible sector and
a lepton asymmetry of opposite sign in the dark sector.
We note that the absence of the decays
$\psi \to \bar{X}+ X'$ and $\phi^* \to L+H$ guarantees
that leptonic asymmetries of equal and opposite sign are generated
in the visible and in the dark sectors.
Indeed, right after the heavy Majorana fermions $N_i$ have decayed completely,
and created the excess of $\psi,\phi$ over $\bar{\psi},\phi^*$,
equal and opposite lepton numbers are already assigned to
the visible sector and the dark sector.
It is clear from the above analysis that there is no violation of lepton number in the entire process of
generating the leptonic asymmetries.
We further note that while sphaleron interactions are active
during the period when the leptogenesis and the genesis of (asymmetric) dark matter occur,
they are not responsible for creating a net $B-L$ number in the visible sector,
though they do play a role in transmuting a part of the lepton number into baryon number in the visible sector.

As will be discussed in Sec.4,
the symmetric component of dark matter would be sufficiently depleted
by annihilating via a $Z'$ pole into standard model particles,
which ensures the asymmetric dark matter to be the dominant component of the current dark matter relic abundance.
One can estimate on general grounds  the mass of the dark particles in this model for the cosmic coincidence to occur.
Since the total $B-L$ in the universe vanishes,
the  $B-L$ number in the visible sector is equal in magnitude and opposite in sign
to the lepton number created in the visible sector right after $N_i$ have completely decayed
(the decay of $N_i$ does not generate any baryon asymmetry),
and thus is equal to the lepton number in the dark sector, i.e.,
\be
   (B-L)_{v}
   = L_{d}\,,
   \label{bl.1}
\ee
where the indices $v,d$ denote the visible sector and the dark sector respectively.
We are interested in the relative density of particle species at the time when
the sphaleron interactions go out of the thermal equilibrium.
This happens at a temperature of $\sim 100$~GeV
which lies below the top mass
so that the top quark would have already decoupled and
no longer participates in the thermal bath.
After the decoupling of the sphaleron interactions $B$ and $L$ are separately conserved
and correspond to the $B$ and $L$ seen today.
An analysis of the chemical potentials~\cite{Harvey:1990qw,Feng:2012jn} allows us to compute the
current value of $B$ in term of $(B-L)_{v}$ so that
\be
\frac{B^f}{(B-L)_{v}} = \frac{30}{97}\,,
\label{bl.2}
\ee
where $B^f$ denotes the final (and currently observed) value of the  baryon number density.
Assuming that $X$ and $X'$ have the same mass,
and using \cref{0.2,bl.1,bl.2} we obtain the mass of the dark particles
\begin{equation}
m_X = m_{X'} \approx 0.85~{\rm GeV}\,.
\end{equation}

\begin{figure}[t!]
\begin{center}
\includegraphics[scale=0.55]{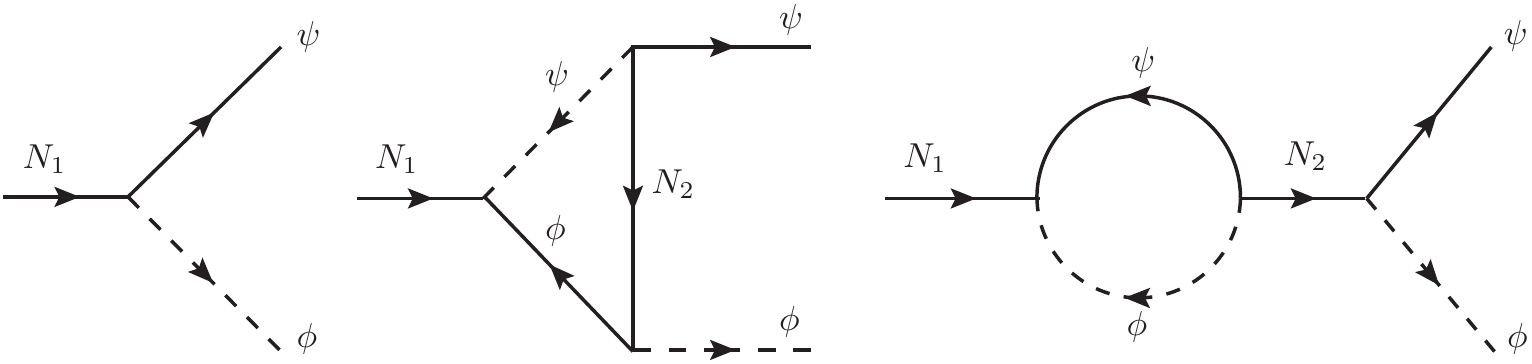}
 \caption{An exhibition of the generation of asymmetry in $\psi,\phi$ over
  their anti-particles $\bar\psi,\phi^*$ from the decay of the  Majorana field $N_1$.
  The lepton number is conserved in these processes.}
\label{FDG1}
\end{center}
\end{figure}

We turn now to the detail of the generation of
the asymmetry between $\psi,\phi$ and $\bar{\psi}, \phi^*$.
We assume there are two Majorana fields $N_1$ and $N_2$
with $N_2$ mass $M_2$ being much larger than the $N_1$ mass $M_1$, i.e., $M_2 \gg M_1$.
Thus the generation of the asymmetry is mostly
through the decay of $N_1$.
The diagrams that contribute to it are shown
in~\cref{FDG1} where the Majorana particles  $N_i$ decay into the Dirac fermion $\psi$
and the complex scalar $\phi$ with $\psi$ and $\phi$
carrying opposite lepton numbers while the Majorana fields $N_i$ carry no lepton number.
As is well-known one needs an interference of the tree and the loop diagrams to create the asymmetry.
The loop diagrams consist of  a vertex diagram and  a wave function diagram
as shown in~\cref{FDG1}.
The excess of  $\psi,\phi$ over $\bar{\psi}, \phi^*$ is given by\footnote{
This calculation is similar to the calculation done in~\cite{Covi:1996wh} for leptogenesis
where the heavy Majorana fields $N_i$ decay to $L,H$ and their anti-particles.
The difference here is that for leptogenesis,
the wave contribution has two diagrams due to $L,H$ being $SU(2)$ doublets;
whereas for our case there is only one diagram for the wave contribution.}
\begin{align}
\epsilon
& =\frac{\Gamma(N_{1}\to \psi\phi)-\Gamma(N_{1}\to\bar\psi\phi^*)}
{\Gamma(N_{1}\to \psi\phi)+\Gamma(N_{1}\to\bar\psi\phi^*)} \nonumber \\
&\simeq
-\frac{1}{8\pi}\frac{{\rm Im}(\lambda_{1}^{2}\lambda_{2}^{*2})}{|\lambda_{1}|^{2}}\frac{M_{1}}{M_{2}}
\,,
\label{ASM}
\end{align}
where we have included both the vertex contribution and the wave contribution.
Since the dark sector does not communicate with the visible sector,
$(B-L)_v$ is equal in magnitude and
opposite in sign to the
 lepton number generated in the visible sector,
\begin{equation}
(B-L)_v = -L_v = -\frac{3}{4} \frac{\kappa\,\epsilon\, \zeta(3) g_N T^3}{\pi^2}\,,
\end{equation}
where  $\zeta(3)\sim 1.202$, $g_N = 2$ for the Majorana field $N_1$,
$\kappa$ is the washout factor~\cite{Buchmuller:2002rq}
due to inverse processes
$\psi+\phi \to N_1,\bar{\psi}+\phi^* \to N_1$
and we assume  $\kappa \sim 0.1$.
Using \cref{bl.2}, one could further link the current baryon number to
the excess of $\psi,\phi$ over $\bar{\psi}, \phi^*$ as
\begin{equation}
\frac{B^f}{s} =\frac{30}{97} \frac{(B-L)_v}{s}
=- \frac{30}{97}\frac{135\zeta(3)}{4\pi^4} \frac{\kappa\, \epsilon}{g_s}\,,
\label{bl.5}
\end{equation}
where the entropy density $s= 2\pi^2 g_s T^3 / 45$ and $g_s \approx 100$
is the entropy degrees of freedom at $T \sim 100$~GeV when the sphaleron interactions decouple.
Using the current astrophysical constraint given in~\cref{0.1}
and  \cref{bl.5} we estimate  $| \epsilon | \sim 10^{-6}$.
\\

{\it Sec.3. Supersymmetric Model\label{sec3}:}
For the supersymmetric case we choose the following set of fields:
 $(\hat{N}_i (i\geq 2), \hat Y, \hat Y', \hat X, \hat X^c, \hat X', \hat X^{\prime c})$ where $\hat{\phantom{a}}$ denotes superfields,
and their lepton numbers
and $U(1)_x$ charges are summarized in \cref{tab1}.
From the table it is clear that $U(1)_x$ is anomaly free and can be gauged.
\begin{table}[h]
\begin{center}
\begin{tabular}{|c|c|c|c|c|c|c|c|}
\hline
 & $\hat{N}_i$ & $\hat Y$ & $\hat Y'$ & $\hat X$ & $\hat X^{c}$ & $\hat X'$ & $\hat X^{\prime c}$\tabularnewline
\hline
$L$ & $0$ & $-1$ & $+1$ & $-\frac{1}{2}$ & $+\frac{1}{2}$ & $-\frac{1}{2}$ & $+\frac{1}{2}$\tabularnewline
\hline
$U(1)_{x}$ & $0$ & $0$ & $0$ & $+1$ & $-1$ & $-1$ & $+1$\tabularnewline
\hline
\end{tabular}
\end{center}
\caption{Lepton numbers and $U(1)_x$ charges of the superfields that enter in the generation of leptonic asymmetries
                  for a gauged $U(1)_x$ model.}
\label{tab1}
\end{table}

For these superfields we assume a superpotential of the following form
which conserve both the lepton number and the $U(1)_x$ gauge symmetry:
\be
 W= W_Y+ W_m\ ,
 \label{w.1}
\ee
 where $W_Y$ contains  the Yukawa couplings
\be
  W_Y=  \lambda_i \hat N_i \hat Y \hat Y'+   \beta  \hat Y \hat L \hat H   +   \beta' \hat Y \hat X^c \hat X^{\prime c} + \gamma \hat Y' \hat X \hat X' \,,
\label{w.2}
\ee
and $W_m$ contains the mass terms
\be
  W_m=   M_i \hat N_i \hat N_i + M_Y \hat Y \hat Y' + m_X \hat X \hat X^c + m_{X'} \hat X' \hat X^{\prime c}\,.
\label{w.3}
\ee
For the supersymmetric model, a possible candidate for $U(1)_x$ is $U(1)_{B-L}$
if one includes three right-handed neutrinos to the particle spectrum.
Along with the anomaly free spectrum of \cref{tab1},
one can then gauge $U(1)_{B-L}$. In this case $\hat Y,\hat Y'$  along with the
dark matter fields $\hat X, \hat X^c, \hat X', \hat X^{\prime c}$ will all  carry $U(1)_{B-L}$
charges as shown in \cref{tab2}.
And of course, the minimal supersymmetric standard model matter fields
also carry $U(1)_{B-L}$ quantum numbers.
In this case we require that  all the fundamental interactions conserve the lepton number
and the $U(1)_{B-L}$ gauge symmetry,
and the superpotentials of \cref{w.2,w.3} remain unchanged.
This model has the very interesting feature in that
we have a gauged $U(1)_{B-L}$ which leads to a conserved $B-L$
with the initial condition $B-L=0$ in the universe.
\begin{table}[h]
\begin{center}
\begin{tabular}{|c|c|c|c|c|c|c|c|}
\hline
 & $\hat{N}_i$ & $\hat Y$ & $\hat Y'$ & $\hat X$ & $\hat X^{c}$ & $\hat X'$ & $\hat X^{\prime c}$\tabularnewline
\hline
$L$ & $0$ & $-1$ & $+1$ & $-\frac{1}{2}$ & $+\frac{1}{2}$ & $-\frac{1}{2}$ & $+\frac{1}{2}$\tabularnewline
\hline
$U(1)_{B-L}$ & $0$ & $+1$ & $-1$ & $+\frac{1}{2}$ & $-\frac{1}{2}$ & $+\frac{1}{2}$ & $-\frac{1}{2}$\tabularnewline
\hline
\end{tabular}
\end{center}
\caption{Lepton numbers and $U(1)_{B-L}$ charges of the superfields that enter in the generation of leptonic asymmetries
                  for a gauged $U(1)_{B-L}$ model.}
\label{tab2}
\end{table}

All the following discussions in this section apply to both of the above two models.
As in the non-supersymmetric case, for the generation of the asymmetry, we assume $\lambda_i$ to be complex,
and $\beta,\beta',\gamma$ are assumed to be real.\footnote{
The interactions $\hat N_i \hat X \hat X^c$ and $\hat N_i \hat X' \hat X^{\prime c}$ could exist.
However, the inclusion of these two interactions will not change our discussion.
This is so because $(\hat X,\hat X^c)$ and $(\hat X',\hat X^{\prime c})$ carry opposite lepton numbers,
and thus there will be no net lepton number generated in the dark sector through $\hat N_i$ decay from these interactions.
Here we assume $\hat N_i$ would mostly decay into $\hat Y,\hat Y'$.}
Again as in the non-supersymmetric case
we assume $i=2$ and assume the $\hat N_2$ mass $M_2$ to be much larger
than the $\hat N_1$ mass $M_1$,
and $M_Y \gg m_X + m_{X'}$.
Again, $M_Y$ could lie in a broad range from order of TeV to much higher scales.
From the interactions of \cref{w.2} we see that
$\hat Y'$ decays exclusively into the dark sector so that $\hat{Y}' \to \hat X+\hat X'$
while $\hat Y$ could decays into the visible sector as well as dark sector particles.
However, with the assumption $|\beta'| \ll | \beta| $,  $\hat{Y}$ will decay
dominantly into visible sector particles, i.e.,  $\hat{Y} \to \hat L+\hat H$.

\begin{figure}[t!]
\begin{center}
\includegraphics[scale=0.55]{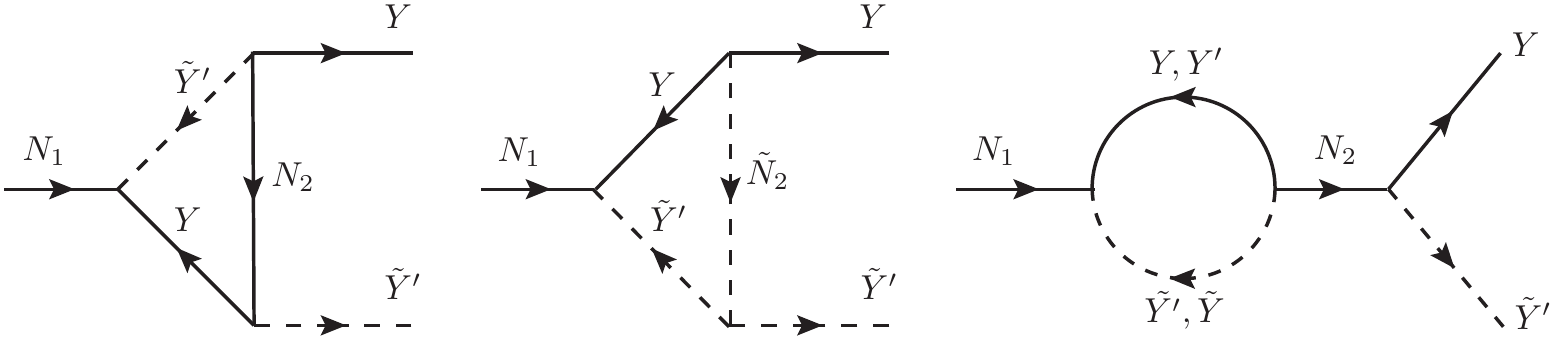}
\caption{Loop diagrams which are responsible for the
  genesis of asymmetry from the decay of $N_1$ to $Y\tilde Y'$.
  There are similar diagrams for the decay of $N_1$ to $\tilde Y Y'$,
  and for the decay of $\tilde N_1$ to $Y Y'$ and to $\tilde{Y} \tilde{Y}'$.
  The  lepton number is conserved in these processes.}
\label{FDG2}
\end{center}
\end{figure}

As in the non-supersymmetric case the asymmetries in $\hat Y$ and  in $\hat Y'$ are generated via
the interference of the tree level amplitudes with the loop diagrams as shown in \cref{FDG2}.
The excess of $\hat{Y},\hat{Y}'$ over their anti-particles $\overline{\hat{Y}},\overline{\hat{Y}'}$
is given by a sum of several $\epsilon$'s,
where these $\epsilon$'s are defined by the final decaying products.
For example, one of these $\epsilon$'s is defined by
\begin{align}
\epsilon_{Y\tilde{Y}'} & =\frac{\Gamma(N_{1}\to Y\tilde{Y}')-\Gamma(N_{1}\to\bar{Y}\tilde{Y}'^*)}{\Gamma(N_{1}\to Y\tilde{Y}')+\Gamma(N_{1}\to\bar{Y}\tilde{Y}'^*)}\,.
\end{align}
Similarly one could define $\epsilon_{\tilde{Y}Y'}$ to parameterize the excess of
$\tilde{Y} Y'$ over $\tilde{Y}^* \bar{Y}'$ decays from $N_1$;
$\epsilon_{Y Y'}$ for the excess of $Y Y'$ over $\bar{Y} \bar{Y}'$ decays from $\tilde{N}_1$;
and $\epsilon_{\tilde{Y} \tilde{Y}'}$ for the excess of $\tilde{Y} \tilde{Y}'$ over $\tilde{Y}^* \tilde{Y}^{\prime *}$
decays from $\tilde{N}_1$.
Similar to the non-supersymmetric case the total asymmetry is a sum of the asymmetries arising
from the interference of the tree diagram with the vertex diagrams
and with the wave function diagram.
An analysis~\cite{Covi:1996wh,Feng:2013wn} of the
asymmetries gives the following relation
\begin{equation}
\epsilon_{Y\tilde{Y}'}=\epsilon_{\tilde{Y}Y'}=\epsilon_{YY'}=\epsilon_{\tilde{Y}\tilde{Y}'}\equiv\varepsilon\, ,
\end{equation}
and in the limit $M_2 \gg M_1$ we obtain~\cite{Feng:2013wn}
\begin{equation}
\varepsilon\simeq -\frac{1}{4\pi}\frac{{\rm Im}(\lambda_{1}^{2}\lambda_{2}^{*2})}{|\lambda_{1}|^{2}}\frac{M_{1}}{M_{2}}\,.
\label{ASUSY}
\end{equation}
The difference between the front factor in \cref{ASUSY} and the front factor in \cref{ASM} is due to the fact
that there are two vertex diagrams for the supersymmetric case
(see \cref{FDG2}) compared to just one vertex diagram for the non-supersymmetric case (see \cref{FDG1}).
Thus the total excess of $\hat{Y}$ over $\overline{\hat{Y}}$,
i.e., $Y,\tilde{Y}$ over $\bar{Y},\tilde{Y}^{*}$
generated by the decay of $\hat{N}_1$ is given by:
\beqn
\Delta n_Y \equiv
n_{\hat{Y}}- n_{\overline{\hat{Y}}}\,,
\eeqn
where $\Delta n_Y$ is computed to be
\beqn
\Delta n_Y
= \big[ \frac{3}{4}(\epsilon_{Y\tilde{Y}'}+\epsilon_{\tilde{Y}Y'}) +(\epsilon_{YY'}+\epsilon_{\tilde{Y}\tilde{Y}'})
\big]
\frac{\kappa\, \zeta(3)g_N T^3 }{\pi^2}\,.\nonumber
\label{DMAsy}
\eeqn
Here the factor of $\frac{3}{4}$ is for $N_1$ and a factor of  1 for $\tilde N_1$,
and again $\kappa$ is a washout factor which we assume to be 0.1.
The excess of $\hat{Y}, \hat{Y}'$ then gives rise to an
equal but opposite lepton number to the visible sector and to the dark sector.
Thus we obtain the $(B-L)$-number density in the visible sector to be
\be
(B-L)_v \approx {2\, \kappa\, \varepsilon}/{g_s}\,,
\ee
where again $g_s \approx 100$
is the entropy degrees of freedom at $T \sim 100$~GeV when the sphaleron interactions decouple.
Similar to the discussion in the non-supersymmetric case,
we estimate $|\varepsilon| \sim 10^{-6}$.

The analysis of the dark matter mass in the supersymmetric model
is also very similar to the one in the non-supersymmetric case, and \cref{bl.1} still holds.
The computation of $B^f$ will be identical to the non-supersymmetric
case and \cref{bl.2} also holds.
This is so because the sleptons and squarks have already decayed into standard model particles
and the memory of them is lost by the time sphaleron interactions go out of thermal equilibrium after which $B$ and $L$ are separately conserved.
The modification that will occur is due to the presence of additional
fields $\hat{X}^c, \hat{X}^{\prime c}$ and both the bosonic and fermionic components of the superfields should be
considered in the analysis. However, the total lepton number will not be affected by
the number of fields.
Assuming the bosonic and the fermionic fields
$\hat X, \hat X', \hat X^c, \hat X^{\prime c}$ all have the same mass, and
again using \cref{0.2,bl.1,bl.2} we obtain
\begin{equation}
m_{X} = m_{X'} = 0.85~{\rm GeV}\,.
\end{equation}
The $U(1)_x$ gaugino $\lambda_x$  is given
a soft mass $\mathcal{L}_{\lambda_x} $ = $ - m_{\lambda} \bar \lambda_x \lambda_x$.
Assuming $m_{\lambda} > m_X + m_{\tilde X}$,
the gaugino $\lambda_x$ can decay into $X\tilde X$ or $X' \tilde X'$, etc, via the supersymmetric interaction
\begin{equation}
\mathcal{L} \sim \bar{\lambda}_x X \tilde X+ \bar{\lambda}_x  X' \tilde X'
+ \bar{\lambda}_x  X^c \tilde X^c + \bar{\lambda}_x  X^{\prime c} \tilde X^{\prime c} +  h.c.\,.
\end{equation}
Thus the gaugino $\lambda_x$ decays into dark particles and is removed from the
low energy  spectrum.
\\

{\it Sec.4. Phenomenology: \label{sec4}}
We discuss now phenomenological implications of the model.
An interesting implication of our model arises in the neutrino sector.
Here we add three families of right-handed neutrinos.
Now we also assume the coupling $\beta$ is family-dependent,
i.e., $\beta \to \beta_i$ where $i=1,2,3$ correspond to $e,\mu,\tau$, c.f., \cref{1.1,w.2}.
The terms which will contribute to neutrino masses read
\begin{equation}
\mathcal{L}_m =
\beta_i \bar \psi_R L_i H
+  \beta_{ij}'' \bar \nu_{iR} L_j H
+ \mu_i' \bar \nu_{iR} \psi_L
+ h.c.\,.
\end{equation}
After spontaneous breaking of the electroweak symmetry, the full mass terms recast into
\begin{equation}
\mathcal{L}_m = \vec{v}_R^T\,\mathcal{M} \,\vec{v}_L + h.c.\,,
\label{mass}
\end{equation}
where we have defined
\begin{gather}
\vec{v}_R^T = \left(\bar{\nu}_{R}^{e},\bar{\nu}_{R}^{\mu},\bar{\nu}_{R}^{\tau},\bar{\psi}_{R}\right)\,, \\
\vec{v}_L^T = \left(\nu_{L}^{e},\nu_{L}^{\mu},\nu_{L}^{\tau},\psi_{L}\right)\,,\\
\mathcal{M} = \left(\begin{array}{cccc}
m_{ee}^\nu & m_{e\mu}^\nu & m_{e\tau}^\nu & \mu_{1}'\\
m_{e\mu}^\nu & m_{\mu\mu}^\nu & m_{\mu\tau}^\nu & \mu_{2}'\\
m_{e\tau}^\nu & m_{\mu\tau}^\nu & m_{\tau\tau}^\nu & \mu_{3}'\\
\mu_1 & \mu_2 & \mu_3 & m_1
\end{array}\right)\,,
\end{gather}
and
\begin{gather}
\mu_i=\frac{1}{\sqrt 2}\beta_i v\,,
\quad m_{ij}^\nu= \frac{1}{\sqrt 2} \beta_{ij}'' v\,,
\label{1.9aa}
\end{gather}
and where $v$ is the VEV of Higgs.
In matrix $\mathcal{M}$, $m_1$ is much larger than all the other entries.

A diagonalization of matrix $\mathcal M$ gives four Dirac fermions in the mass eigenbasis:
three of which correspond to the three neutrinos,
while the fourth one is mostly constituted by $\psi$ which is much heavier.
However, a fine-tuning is needed
to get the light neutrino masses in the experimental range.
$\mathcal M$ can be diagonalized by using a biunitary transformation so that
\be
 V^{\dagger} \mathcal{M} U = \mathcal{M}_D\,.
\label{1.9c}
\ee
Thus the left-handed neutrino states transform as
\be
\nu_{iL} = \sum_{a=1}^4 U_{ia} \nu_{aL}'\,,
\label{1.9d}
\ee
where $\nu_{aL}'$ are in the mass diagonal basis. \cref{1.9d} implies that,
for example, the partial decay widths of the
$W$ and $Z$ bosons will be modified so that
\begin{align}
\Gamma(W \to  \ell_i \bar{\nu}_i) &= \Gamma(W \to  \ell_i \bar{\nu}_i)_{\rm SM}  (1-|U_{i4}|^2)\,,\\
\Gamma(Z \to \nu_i \bar{\nu}_i) &= \Gamma(Z \to \nu_i \bar{\nu}_i)_{\rm SM}  (1-|U_{i4}|^2)^2\,.
\label{1.9f}
\end{align}

Now the low energy electroweak data is in excellent agreement with the standard model and
thus the new physics can be accommodated only within the error bars. Here we use
the data on the hidden decays of the $Z$ boson~\cite{Beringer:1900zz},
which in the standard model are neutrinos,
to constrain $U_{i4}$, i.e.,
\begin{equation}
    \Gamma(Z\to \nu \bar \nu)=  (499\pm 1.5)~\MeV\,.
\label{1.17}
\end{equation}
Using  \cref{1.17} and assuming the correction $U_{i4}$ is uniform across generations
we get an upper limit on $U_{i4}$ of
\be
   |U_{i4}| \lesssim 4\times 10^{-2}\,.
   \label{1.17a}
\ee
The presence of a sizable $U_{i4}$ will also affect other electroweak processes where neutrinos appear.
Thus more accurate measurements in the electroweak sector in the future, for example, at the ILC could reveal
the presence of a non-negligible value of $U_{i4}$. This would provide a possible test of the model.

Next we demonstrate that a sizable $U_{i4}$  can be obtained from \cref{mass}
consistent with small neutrino masses.
We first consider an example of one generation of neutrino  (say the third generation) mixing with the $\psi$ field.
For this case we have
\begin{equation}
\mathcal{L}_{m}^{(3)}=\left(\bar{\nu}_{3R},\bar{\psi}_{R}\right)\left(\begin{array}{cc}
m_{\tau\tau}^\nu & \mu_3'\\
\mu_3 & m_{1}
\end{array}\right)\left(\begin{array}{c}
\nu_{3L}\\
\psi_{L}
\end{array}\right) + h.c.\,.
\label{1.17aa}
\end{equation}
With the inputs $m_{\tau\tau}=10^{-12},\mu_3'=10^{-9}, \mu_3=10, m_{1}=1000$ (all masses in GeV),
we obtain the mass eigenvalue of the neutrino to be around $10^{-2}$~eV,
$U_{34}\sim0.01$ consistent with the constraint of \cref{1.17a}.
The $2\times 2$ matrix analysis above uses a lopsided matrix in \cref{1.17aa}.
An analysis of the lopsided $4\times 4$ case is more elaborate and for that reason we do not give an extended analysis
of this case here but we expect that a sizable $U_{i4}$ can be generated in that case as well.

We discuss now another sector of the parameter space of \cref{mass}. Here we assume a symmetrical
form for the neutrino mass terms so that
\begin{equation}
\mathcal{L}_{m}^\nu=
\vec{v}_R^T
\left(\begin{array}{cccc}
m_{\nu_{e}} & 0 & 0 & \mu_{1}'\\
0 & m_{\nu_{\mu}} & 0 & \mu_{2}'\\
0 & 0 & m_{\nu_{\tau}} & \mu_{3}'\\
\mu_1 & \mu_2 & \mu_3 & m_{1}
\end{array}\right)
\vec{v}_L
+h.c.\,,
\label{neutrino-matrix2}
\end{equation}
and we further assume
\begin{equation}
\mu_1 = \mu_1'\,,\quad \mu_2= \mu_2'\,,\quad \mu_3 = \mu_3'\,.
\label{1.210}
\end{equation}
The matrix of \cref{neutrino-matrix2} contains no direct mixings among the neutrino flavor states.
However, we will see that their mixings with the field $\psi$ automatically lead us to neutrino flavor mixings.
To exhibit this result we  diagonalize the matrix of \cref{neutrino-matrix2} by an orthogonal
transformation.

By setting   $m_{\nu_{e}}=10^{-11},m_{\nu_{\mu}}=1.7\times10^{-10},m_{\nu_{\tau}}=2\times10^{-9},m_1=2000,
\mu_1=3.6\times10^{-5},\mu_2=8.9\times10^{-5},\mu_3=5.9\times10^{-4}$
(all masses in GeV) the three neutrino masses in the mass diagonal basis are calculated to be
\begin{align}
m_{3} & \approx 4.8 \times10^{-2}~{\rm eV} ,\\
m_{2} & \approx 1.2 \times10^{-2}~{\rm eV} ,\\
m_{1} & \approx 4.2 \times10^{-3}~{\rm eV}  ,
\label{1.211}
\end{align}
which  produce the normal hierarchy of neutrino masses~\cite{Beringer:1900zz}
and the mass eigenvalue of the heavy
field $\psi$ is still approximately $m_{1}$.
For the neutrino mixings we obtain
\be
\sin^2\theta_{12} \approx 0.30\,, \ \sin^2\theta_{23} \approx 0.36\,,\ \sin^2\theta_{13} \approx 0.024\,,
\label{1.212}
\ee
while the experimental values are~\cite{Beringer:1900zz}
\begin{gather}
\sin^2\theta_{12} = 0.307^{+0.018}_{-0.016}\,,\quad
\sin^2\theta_{23} = 0.386^{+0.024}_{-0.021}\,,\nonumber \\
\sin^2\theta_{13}= 0.0244^{+0.0023}_{-0.0025}\,.
\label{1.213}
\end{gather}

We see that our analysis of \cref{1.212} is in good accord with the experimental determination of
the mixing angles as given in \cref{1.213}. Specifically the model is consistent with the result
from the Daya Bay neutrino reactor experiment~\cite{An:2012eh}
of $\theta_{13} \sim 9^\circ$.  Thus it is very interesting that the model provides an explanation
of the neutrino mixings at a fundamental level, in that the neutrino mixings arise as a consequence
of the interaction of the neutrinos with the primordial Dirac field $\psi$ which enters in leptogenesis
which points to the cosmological origin of neutrino mixings.

\begin{figure}[t!]
  \begin{center}
  \includegraphics[scale=0.36]{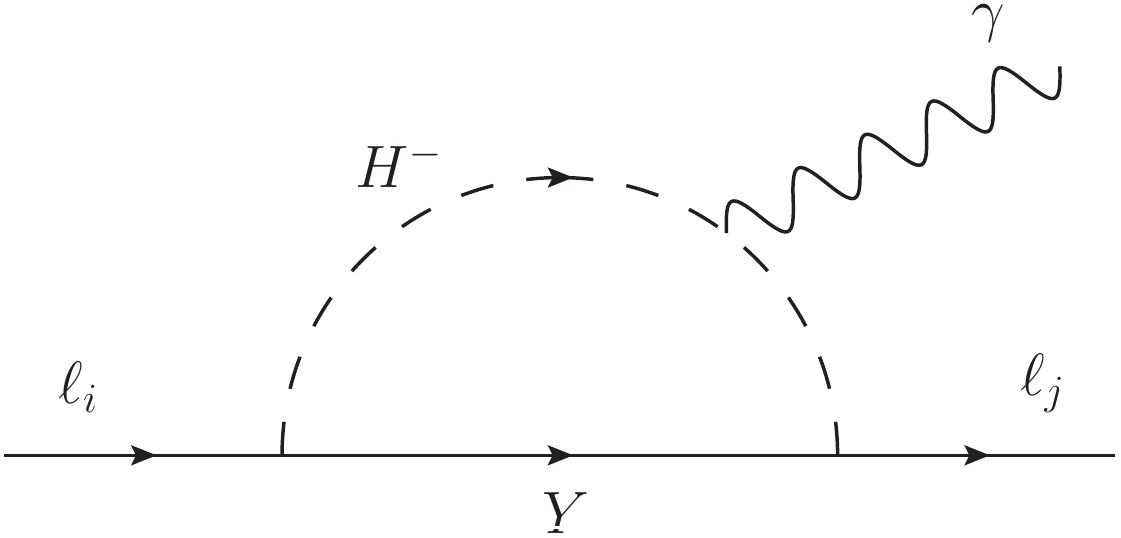}
  \caption{Flavor changing processes $\ell_i \to \ell_j \gamma$ via the charged Higgs and $Y$ loop.}
  \label{muega}
  \end{center}
\end{figure}

Other implications of the model involve flavor changing processes.
For the supersymmetric model of \cref{w.2},
after spontaneous breaking one has interactions of the
charged Higgs $H^+$ with charged leptons and $Y$:
\be
 {\cal L}_{H\ell\psi}= \beta_i \bar Y \ell_i H^+ +  h.c.\,,
\ee
where $\ell_i$ denotes the charged leptons.
Such interactions will give rise to $\ell_i \to \ell_j \gamma$ processes,
where a charged lepton $\ell_i$ converts into a charged lepton $\ell_j$ via exchange of $Y$ while a photon
is emitted by the charged Higgs inside the loop,  see \cref{muega}.
Assuming $m_Y^2 \gg m_{H^+}^2$,
we obtain the decay rate of the flavor changing process $\ell_i \to \ell_j \gamma$ to be\footnote{
Loops which involve the Higgsinos and $\tilde Y$ also contribute to $\ell_i \to \ell_j \gamma$ process.
A computation shows that these loops are suppressed by a factor of $m_{\tilde H}^2 / m_{\tilde Y}^2$ compare to
the charged Higgs and $Y$ loop. For the case $m_{\tilde Y}^2 \gg m_{\tilde H}^2$, we could omit these contributions.}
\begin{equation}
\d \Gamma_{\ell_i \to \ell_j \gamma} = \frac{\alpha_{\rm em} (\beta_i \beta_j)^2}{(16 \pi^2)^2} \frac{m_i^3}{M_Y^2}\,,
\label{DW}
\end{equation}
where $m_i$ is the mass of the decaying charged lepton and we have used $m_i \gg m_j$.
The current experimental upper bounds on the branching ratio of such flavor changing processes read~\cite{Adam:2011ch,Aubert:2009ag}
\begin{align}
\label{br.1}
\mathcal{B}r(\mu \to e\gamma) &\lesssim 2.4\times 10^{-12}\,, \\
\label{br.2}
\mathcal{B}r(\tau \to e\gamma) &\lesssim 3.3\times 10^{-8}\,, \\
\label{br.3}
\mathcal{B}r(\tau \to \mu \gamma) &\lesssim 4.4\times 10^{-8}\,.
\end{align}
Using the mean lifetimes for  $\mu$ and $\tau$~\cite{Beringer:1900zz},
the branching ratios \cref{br.1,br.2,br.3}
 and \cref{DW}, and $M_Y \sim 1~{\rm TeV}$ we obtain
\begin{align}
\beta_1 \sim \beta_2 \lesssim 3\times 10^{-3}\,.
\end{align}
Once $\beta_1,\beta_2$ are fixed, one can estimate $\beta_3$ by
\begin{equation}
\beta_3 \lesssim 2 \times 10^{-4} / \beta_1\,.
\end{equation}
One can expect observable effects in these flavor changing processes
in future experiments with improved sensitivities.
And at the same time, we see that with these constraints,
$\mu_3$ could be of $\mathcal{O}(10)$~GeV, c.f., \cref{1.9aa},
thus one would also expect to see the effect we discussed at \cref{1.17aa}.

Next we discuss the phenomenological implications of the model in the dark sector.
An important issue concerns the dissipation of thermally produced dark matter.
To dissipate the symmetric component of dark matter  we use  the fact that dark matter is charged
under  the gauged group $U(1)_x$ or $U(1)_{B-L}$.
We assume that the  $U(1)$ gauge boson gains mass
via the Stueckelberg mechanism~\cite{Kors:2004dx}.\footnote{
An alternative way of depleting the symmetric component of the dark matter is
assuming the $U(1)_x$ gauge boson to be massless (dark photon).
Then the symmetric component of the dark matter could sufficiently annihilate
into the $U(1)_x$ dark photons and become radiation in the early universe.
As shown in~\cite{Blennow:2012de},
the constraints on the number of extra effective neutrino species, $\Delta N_{\rm eff}$,
can be satisfied for a large class of asymmetric dark matter models.}
For the gauged $U(1)_x$ model, one could assume a kinetic mixing
of the $U(1)_x$ gauge boson with the $U(1)_Y$ gauge boson~\cite{Holdom:1985ag}.
This mechanism allows one to dissipate the symmetric component of dark matter
which can annihilate into the standard model particles via the $Z-Z'$ mixing.
The analysis here is very similar to the ones discussed in~\cite{Feng:2013wn}.

The $Z'$ gauge boson can make a contribution to the anomalous magnetic moment of the muon.
At the one loop order one finds
 \be
 \Delta a_{\mu} \simeq \delta^2 \frac{g_Y^2 m_\mu^2}{48 \pi^2 M_{Z'}^2}\,,
 \ee
where $\delta$ is the coupling of the $Z'$ with  matter current, i.e.,
${\cal L}_{\rm int}^{Z'}= \delta\, Z'_\mu J^\mu $.  A value of $\delta \sim 10^{-3}$ and a $Z'$ mass of
order of a few GeV gives $\Delta a_{\mu}$ significantly below the current experimental limit on the
deviation of $a_\mu$ from the standard model value of  $\Delta (a_{\mu}) < 3\times 10^{-9}$~\cite{Beringer:1900zz}.
At the same time  $\delta$ and $M_{Z'}$ satisfy the LEP~II constraint of~\cite{LEP:2003aa} that
$M_{Z'}/g_{Z' \bar \ell \ell} > 6$~TeV.

As discussed earlier, fields $X,X'$ (non-supersymmetric case) or
$\hat{X}, \hat{X}^c, \hat{X}', \hat{X}^{\prime c}$ (supersymmetric case) constitute the dark matter
which are all light with masses $\mathcal{O}(1)$~GeV.
Since the coupling between $Z'$
and standard model particles can only be $\sim10^{-3}$ because of experimental constraints,
a sufficient depletion of  the symmetric component of dark matter
(up to or less than  $10\%$ of the total dark matter relic density),
requires a Breit-Wigner enhancement, so that the $Z'$ mass is around twice the dark matter mass.
It is seen that with a kinetic mixing parameter $\delta\sim0.001$~\cite{Mambrini:2010dq},
for a dark matter mass of $\sim1$~GeV, a $Z'$  mass of $\sim 3$~GeV does allow
the symmetric component of dark matter to be depleted down to less than $10\%$ of the total dark matter relic density.
Thus the current dark matter would be constituted of up to $90\%$ or more of the light asymmetric dark matter.

Such dark matter can scatter from quarks within a nucleon through the t-channel exchange of the $Z'$ boson.
The spin-independent dark matter-nucleon (target-independent) cross
section can be approximately written as~\cite{Morrissey:2009ur,Frandsen:2011cg}
\begin{equation}
\sigma_{{\rm SI}}\sim\frac{4}{\pi}\frac{\delta^{2} g_{x}^{2} g_{Y}^{2} \cos^{4}\theta_{W}\mu_{n}^{2}}{m_{Z'}^{4}}\,,
\end{equation}
where $\mu_{n}$ is the dark matter-nucleon reduced mass.
Using the parameters discussed above we find $\sigma_{{\rm SI}}\sim10^{-37}~{\rm cm}^{2}$,
which is just on the edge of sensitivity of the CRESST~I experiment~\cite{Angloher:2002in}.
Thus improved experiments in the future in the low dark matter mass region
with better sensitivities should be able to test the model.

For the supersymmetric gauged $U(1)_{B-L}$ model, without using the kinetic mixing mechanism,
one could use the $U(1)_{B-L}$ gauge boson to dissipate the symmetric component of dark matter.
As discussed in~\cite{Feng:2012jn}, the mass of the $U(1)_{B-L}$ gauge boson can lie in a few GeV range
and be consistent with the LEP~II constraints and with the UA2 cross section bounds~\cite{Alitti:1993pn}.
The analysis of~\cite{Feng:2012jn} also shows that
the symmetric component of dark matter can be sufficiently depleted.

There could be also indirect hints for the existence of the asymmetric dark matter.
For example, assume that dark matter consists of both an asymmetric component
which is the dominant one ($\gtrsim 90\%$) and a subdominant component ($\lesssim 10\%$) which is WIMP like.
A detailed analysis shows that the subdominant component could still be detected~\cite{Feng:2012jn}.
On the other hand the WIMP model would not constitute
the entire relic density which would require the asymmetric
dark matter to make up the deficit.
This could provide an indirect evidence for asymmetric dark matter
if WIMPs were observed in direct detection but a detailed theory model shows
a large deficit in its contribution to the relic density.

Thus  quite interestingly the  above discussion indicates that
the cogenesis model which relates to  cosmological issue gets directly related to particle physics
experiments  specifically experiments at the intensity frontier~\cite{Hewett:2012ns} and
those related to search for dark matter.

Finally, we compare our work briefly with the work of other authors
specifically the works of \cite{Davoudiasl:2010am,Gu:2010ft}.
There are major differences between our work and theirs
both in the structure of the model as well as in the phenomenological implications.
At the level of the structure of the model
the major difference between our model and the models of \cite{Davoudiasl:2010am,Gu:2010ft} is that
for our model the asymmetries in both the visible and the dark sectors are generated
through the decay of heavy Majorana fields, which do not carry any lepton or baryon numbers,
while for the model of \cite{Davoudiasl:2010am} the asymmetries in both sectors
arise from the decay of heavy Dirac particles which carry baryon number
and for the model of  \cite{Gu:2010ft} the asymmetries arise
from the decay of  heavy complex scalars which carry either baryon or lepton number.

Further, in our model we have two mediator fields ($\psi,\phi$ in the non-supersymmetric case
and $\hat{Y}, \hat{Y}'$ in the supersymmetric case) which subsequently decay into visible or dark sector particles
after they are produced by the decay of the heavy Majorana fields, and this procedure has the advantage
that the experimental data on the asymmetry in the universe does not set a bound on
the mass of the mediator fields or on the couplings of
the mediator fields to the standard model particles.
In addition, our model is focused on generating the asymmetry in the leptonic sector,
whereas the work of \cite{Davoudiasl:2010am} focused on generating the asymmetry in the baryonic sector.
Although the work of \cite{Gu:2010ft} also has a model on generating the asymmetry in the  leptonic sector,
that model is very different from ours.

In addition to differences in the theoretical structure of the models,
there are very significant phenomenological
differences between the model  presented here and the works of \cite{Davoudiasl:2010am,Gu:2010ft}.
The phenomenological implications of the works of \cite{Davoudiasl:2010am,Gu:2010ft} have been spelled
out in these works and we do not wish to enumerate them here.
One item, however which we wish to point out it that in the model of \cite{Davoudiasl:2010am}
that dark particles can induce proton decay.
This feature is not shared by our model.
In contrast we have discussed in this work a variety of phenomena arise from the leptonic sector,
which can provide low energy tests of the proposed model.
These include implication of the model for electroweak physics,
neutrino masses and mixings and lepton flavor changing processes.
Further, the mechanism for the dissipation of symmetric component of dark matter in the model is also different.
In particular, we propose that the symmetric component of dark matter can annihilate efficiently into
standard model particles through a very light $Z'$ with mass around twice the dark matter mass~\cite{Feng:2012jn,Feng:2013wn}.
This $Z'$ gauge boson can only couple very weakly to the standard model particles
and thus satisfies all the current experimental constraints.
In summary, both the theoretical structure and the phenomenological implications of the proposed models are very
different from previous works on this topic.\\

{\it Sec.5.   Conclusion:\label{sec5}}
In this work we discussed  models of leptogenesis where the fundamental interactions do not
violate lepton number, and the total $B-L$ in the universe vanishes.
Thus the generation of a net lepton number in the visible universe
is compensated by the generation of an equal amount of anti-lepton number in the
dark sector. Baryogenesis in this class of models occurs via the sphaleron interactions
which convert a part of the lepton number in the visible sector into baryon number  in the visible sector.
Three models are discussed in this work: one non-supersymmetric gauged $U(1)_x$ model,
one supersymmetric  gauged $U(1)_x$ model, and another supersymmetric gauged $U(1)_{B-L}$ model.
A detailed analysis shows that the models can generate the baryon number
density to the entropy density ratio consistent with the observed value. The models also produce
the desired amount of dark matter and provide an explanation of the observed dark matter density
to the baryonic matter density in the universe. Thus the proposed  models provide a possible explanation of the
three cosmology puzzles mentioned in the introduction.
The results from the analysis of these models indicate
that a violation of lepton number in the fundamental interactions is not essential for leptogenesis.
The symmetric component of dark matter in these models is dissipated
via kinetic mixing between $U(1)_x$ and $U(1)_Y$ gauge bosons,
or for the supersymmetric gauged $U(1)_{B-L}$ model through a light $Z'_{B-L}$ gauge boson.
The gauged $U(1)_{B-L}$ model is rather attractive in that the $U(1)_{B-L}$ gauge invariance
requires conservation of $B-L$ and $B-L=0$ provides the most natural initial condition for the
universe.

Phenomenological implications of the models were discussed.
These include implications for electroweak physics and neutrino masses and mixings.
Specifically it is seen that small corrections arise in the electroweak sector which
may be detectable in future high precision machines such as ILC.
Further, it is seen that in the proposed models neutrino interactions
with a primordial Dirac field which enters in leptogenesis naturally lead to neutrino mixings
after spontaneous breaking of the electroweak symmetry.
For the supersymmetric case, our model may lead to
the charged lepton flavor changing processes such as $\ell_i\to \ell_j \gamma$ at the loop level.
Also discussed is the feasibility of the direct detection of dark matter in $\mathcal{O}(1)$~GeV mass range.
The models proposed in this work are significantly different from other cogenesis models
both in theoretical structure as well as in their phenomenological implications.\\

\noindent
{\it Acknowledgments:}
WZF is grateful to Javier Redondo, Leo Stodolsky and Wei Xue for helpful discussions.
The work of PN is supported in part by the U.S. National Science Foundation (NSF) grant
PHY-1314774. WZF is supported by the Alexander von Humboldt Foundation.


\begin{thebibliography}{999}

\bibitem{wmap}
  E.~Komatsu {\it et al.}  [WMAP Collaboration],
  Astrophys.\ J.\ Suppl.\  {\bf 192}, 18 (2011).
  [arXiv:1001.4538 [astro-ph.CO]].

\bibitem{Sakharov:1967dj}
  A.~D.~Sakharov,
  Pisma Zh.\ Eksp.\ Teor.\ Fiz.\  {\bf 5}, 32 (1967)
  [JETP Lett.\  {\bf 5}, 24 (1967)]
  [Sov.\ Phys.\ Usp.\  {\bf 34}, 392 (1991)]
  [Usp.\ Fiz.\ Nauk {\bf 161}, 61 (1991)].

\bibitem{Ade:2013ktc}
  P.~A.~R.~Ade {\it et al.}  [Planck Collaboration],
  [arXiv:1303.5062 [astro-ph.CO]].

\bibitem{AsyDM}
  D.~E.~Kaplan, M.~A.~Luty and K.~M.~Zurek,
  Phys.\ Rev.\ D {\bf 79}, 115016 (2009)
  [arXiv:0901.4117 [hep-ph]].
  %
  T.~Cohen, D.~J.~Phalen, A.~Pierce and K.~M.~Zurek,
  Phys.\ Rev.\ D {\bf 82}, 056001 (2010)
  [arXiv:1005.1655 [hep-ph]].
  %
  M.~L.~Graesser, I.~M.~Shoemaker and L.~Vecchi,
  JHEP {\bf 1110}, 110 (2011)
  [arXiv:1103.2771 [hep-ph]].
  %
  M.~Ibe, S.~Matsumoto and T.~T.~Yanagida,
  Phys.\ Lett.\ B {\bf 708}, 112 (2012)
  [arXiv:1110.5452 [hep-ph]].

\bibitem{Feng:2012jn}
  W.~-Z.~Feng, P.~Nath and G.~Peim,
  Phys.\ Rev.\ D {\bf 85}, 115016 (2012)
  [arXiv:1204.5752 [hep-ph]].

\bibitem{review}
  H.~Davoudiasl and R.~N.~Mohapatra,
  New J.\ Phys.\  {\bf 14}, 095011 (2012)
  [arXiv:1203.1247 [hep-ph]].
%
  K.~Petraki and R.~R.~Volkas,
  Int.\ J.\ Mod.\ Phys.\ A {\bf 28}, 1330028 (2013)
  [arXiv:1305.4939 [hep-ph]].
  K.~M.~Zurek,
  [arXiv:1308.0338 [hep-ph]].

\bibitem{darkogenesis}
  N.~Haba and S.~Matsumoto,
  Prog.\ Theor.\ Phys.\  {\bf 125}, 1311 (2011)
  [arXiv:1008.2487 [hep-ph]].
  M.~R.~Buckley and L.~Randall,
  JHEP {\bf 1109}, 009 (2011).
  [arXiv:1009.0270 [hep-ph]].
  J.~Shelton and K.~M.~Zurek,
  Phys.\ Rev.\ D {\bf 82}, 123512 (2010)
  [arXiv:1008.1997 [hep-ph]].

\bibitem{Feng:2013wn}
  W.~-Z.~Feng, A.~Mazumdar and P.~Nath,
  Phys.\ Rev.\ D {\bf 88}, 036014 (2013)
  [arXiv:1302.0012 [hep-ph]].

\bibitem{cogenesis}
  M.~Y.~.Khlopov and C.~Kouvaris,
  Phys.\ Rev.\ D {\bf 77}, 065002 (2008)
  [arXiv:0710.2189 [astro-ph]].
  %
  P.~-H.~Gu and U.~Sarkar,
  Phys.\ Rev.\ D {\bf 81}, 033001 (2010)
  [arXiv:0909.5463 [hep-ph]].
  %
  H.~An, S.~-L.~Chen, R.~N.~Mohapatra and Y.~Zhang,
  JHEP {\bf 1003}, 124 (2010)
  [arXiv:0911.4463 [hep-ph]].
  %
  E.~J.~Chun,
  Phys.\ Rev.\ D {\bf 83}, 053004 (2011)
  [arXiv:1009.0983 [hep-ph]].
  %
  B.~Dutta and J.~Kumar,
  Phys.\ Lett.\ B {\bf 699}, 364 (2011)
  [arXiv:1012.1341 [hep-ph]].
  %
  A.~Falkowski, J.~T.~Ruderman and T.~Volansky,
  JHEP {\bf 1105}, 106 (2011)
  [arXiv:1101.4936 [hep-ph]].
  %
  E.~J.~Chun,
  JHEP {\bf 1103}, 098 (2011)
  [arXiv:1102.3455 [hep-ph]].
  %
  J.~March-Russell and M.~McCullough,
  JCAP {\bf 1203}, 019 (2012)
  [arXiv:1106.4319 [hep-ph]].
  %
  C.~Arina and N.~Sahu,
  Nucl.\ Phys.\ B {\bf 854}, 666 (2012)
  [arXiv:1108.3967 [hep-ph]].
  K.~Petraki, M.~Trodden and R.~R.~Volkas,
  JCAP {\bf 1202}, 044 (2012)
  [arXiv:1111.4786 [hep-ph]].
  %
  K.~Kamada and M.~Yamaguchi,
  Phys.\ Rev.\ D {\bf 85}, 103530 (2012)
  [arXiv:1201.2636 [hep-ph]].
  %
  C.~Arina, J.~-O.~Gong and N.~Sahu,
  Nucl.\ Phys.\ B {\bf 865}, 430 (2012)
  [arXiv:1206.0009 [hep-ph]].
  %
  H.~Kuismanen and I.~Vilja,
  Phys.\ Rev.\ D {\bf 87}, 015005 (2013)
  [arXiv:1210.4335 [hep-ph]].
  %
  P.~Fileviez Perez and M.~B.~Wise,
  JHEP {\bf 1305}, 094 (2013)
  [arXiv:1303.1452 [hep-ph]].

\bibitem{oscillations}
  M.~R.~Buckley and S.~Profumo,
  Phys.\ Rev.\ Lett.\  {\bf 108}, 011301 (2012)
  [arXiv:1109.2164 [hep-ph]].
  M.~Cirelli, P.~Panci, G.~Servant and G.~Zaharijas,
  JCAP {\bf 1203}, 015 (2012)
  [arXiv:1110.3809 [hep-ph]].

\bibitem{Fukugita:1986hr}
  M.~Fukugita and T.~Yanagida,
  Phys.\ Lett.\ B {\bf 174}, 45 (1986).

\bibitem{Covi:1996wh}
  L.~Covi, E.~Roulet and F.~Vissani,
  Phys.\ Lett.\ B {\bf 384}, 169 (1996)
  [hep-ph/9605319].

\bibitem{Davoudiasl:2010am}
  H.~Davoudiasl, D.~E.~Morrissey, K.~Sigurdson and S.~Tulin,
  Phys.\ Rev.\ Lett.\  {\bf 105}, 211304 (2010)
  [arXiv:1008.2399 [hep-ph]].
A supersymmetric version of the model is discussed in
  N.~Blinov, D.~E.~Morrissey, K.~Sigurdson and S.~Tulin,
  Phys.\ Rev.\ D {\bf 86}, 095021 (2012)
  [arXiv:1206.3304 [hep-ph]].

\bibitem{Gu:2010ft}
  P.~-H.~Gu, M.~Lindner, U.~Sarkar and X.~Zhang,
  Phys.\ Rev.\ D {\bf 83}, 055008 (2011)
  [arXiv:1009.2690 [hep-ph]].

\bibitem{Perez:2013tea}
  P.~F.~Perez and H.~H.~Patel,
  arXiv:1311.6472 [hep-ph].

\bibitem{DLeptogenesis}
  K.~Dick, M.~Lindner, M.~Ratz and D.~Wright,
  Phys.\ Rev.\ Lett.\  {\bf 84}, 4039 (2000)
  [hep-ph/9907562].
  H.~Murayama and A.~Pierce,
  Phys.\ Rev.\ Lett.\  {\bf 89}, 271601 (2002)
  [hep-ph/0206177].

\bibitem{Harvey:1990qw}
  J.~A.~Harvey and M.~S.~Turner,
  Phys.\ Rev.\ D {\bf 42}, 3344 (1990).

\bibitem{Buchmuller:2002rq}
  W.~Buchmuller, P.~Di Bari and M.~Plumacher,
  Nucl.\ Phys.\ B {\bf 643}, 367 (2002)
  [hep-ph/0205349];
  Nucl.\ Phys.\ B {\bf 665}, 445 (2003).
  [hep-ph/0302092].

\bibitem{Beringer:1900zz}
  J.~Beringer {\it et al.}  [Particle Data Group Collaboration],
  Phys.\ Rev.\ D {\bf 86}, 010001 (2012).

\bibitem{An:2012eh}
  F.~P.~An {\it et al.}  [DAYA-BAY Collaboration],
  Phys.\ Rev.\ Lett.\  {\bf 108}, 171803 (2012)
  [arXiv:1203.1669 [hep-ex]].

\bibitem{Adam:2011ch}
  J.~Adam {\it et al.}  [MEG Collaboration],
  Phys.\ Rev.\ Lett.\  {\bf 107}, 171801 (2011)
  [arXiv:1107.5547 [hep-ex]].

\bibitem{Aubert:2009ag}
  B.~Aubert {\it et al.}  [BaBar Collaboration],
  Phys.\ Rev.\ Lett.\  {\bf 104}, 021802 (2010)
  [arXiv:0908.2381 [hep-ex]].

\bibitem{Kors:2004dx}
  B.~Kors and P.~Nath,
  Phys.\ Lett.\ B {\bf 586}, 366 (2004);
  [hep-ph/0402047];
  JHEP {\bf 0412}, 005 (2004);
  [hep-ph/0406167];
  JHEP {\bf 0507}, 069 (2005);
  [hep-ph/0503208].
  D.~Feldman, Z.~Liu and P.~Nath,
  Phys.\ Rev.\ D {\bf 75}, 115001 (2007).
  [hep-ph/0702123].

\bibitem{Blennow:2012de}
  M.~Blennow, E.~Fernandez-Martinez, O.~Mena, J.~Redondo and P.~Serra,
  JCAP {\bf 1207}, 022 (2012)
  [arXiv:1203.5803 [hep-ph]].

\bibitem{Holdom:1985ag}
  B.~Holdom,
  Phys.\ Lett.\ B {\bf 166}, 196 (1986);
  Phys.\ Lett.\ B {\bf 259}, 329 (1991).

\bibitem{LEP:2003aa}
  t.~S.~Electroweak [LEP and ALEPH and DELPHI and L3 and OPAL and LEP Electroweak Working Group and SLD Electroweak Group and SLD Heavy Flavor Group Collaborations],
  [hep-ex/0312023].

\bibitem{Mambrini:2010dq}
  Y.~Mambrini,
  JCAP {\bf 1009}, 022 (2010)
  [arXiv:1006.3318 [hep-ph]].

\bibitem{Morrissey:2009ur}
  D.~E.~Morrissey, D.~Poland and K.~M.~Zurek,
  JHEP {\bf 0907}, 050 (2009)
  [arXiv:0904.2567 [hep-ph]].

\bibitem{Frandsen:2011cg}
  M.~T.~Frandsen, F.~Kahlhoefer, S.~Sarkar and K.~Schmidt-Hoberg,
  JHEP {\bf 1109}, 128 (2011)
  [arXiv:1107.2118 [hep-ph]].

\bibitem{Angloher:2002in}
  G.~Angloher, S.~Cooper, R.~Keeling, H.~Kraus, J.~Marchese, Y.~A.~Ramachers, M.~Bruckmayer and C.~Cozzini {\it et al.},
  Astropart.\ Phys.\  {\bf 18}, 43 (2002).

\bibitem{Alitti:1993pn}
  J.~Alitti {\it et al.}  [UA2 Collaboration],
  Nucl.\ Phys.\ B {\bf 400}, 3 (1993).

\bibitem{Hewett:2012ns}
  J.~L.~Hewett, H.~Weerts, R.~Brock, J.~N.~Butler, B.~C.~K.~Casey, J.~Collar, A.~de Gouvea and R.~Essig {\it et al.},
  arXiv:1205.2671 [hep-ex].



\end{thebibliography}
\end{document}